\documentclass[conference]{IEEEtran}
\IEEEoverridecommandlockouts
\usepackage{cite}
\usepackage{amsmath,amssymb,amsfonts}
\usepackage{algorithmic}
\usepackage{graphicx}
\usepackage{textcomp}
\usepackage{xcolor}
\usepackage{fancyhdr}

\def\BibTeX{{\rm B\kern-.05em{\sc i\kern-.025em b}\kern-.08em
    T\kern-.1667em\lower.7ex\hbox{E}\kern-.125emX}}

\begin{document}

\title{Novel Semi-parametric Tobit Additive Regression Models\\
}

\author{\IEEEauthorblockN{Hailin Huang}
	
\IEEEauthorblockA{\textit{Department of Statistics} \\
\textit{George Washington University}\\
Washington DC \\
hhl1988@gwmail.gwu.edu}\\
}

\maketitle

\begin{abstract}
Regression method has been widely used to explore relationship between dependent and independent variables. In practice, data issues such as censoring and missing data often exist. When the response variable is (fixed) censored, Tobit regression models have been widely employed to explore the relationship between the response variable and covariates. In this paper, we extend conventional parametric Tobit models to a novel semi-parametric regression model by replacing the linear components in Tobit models with nonparametric additive components, which we refer as Tobit additive models, and propose a likelihood based estimation method for Tobit additive models.  
Numerical experiments are conducted to evaluate the finite sample performance. The estimation method works well in finite sample experiments, even when sample size is relative small.
\end{abstract}


\section{Introduction}
Regression analysis has been one of the most important statistical techniques in data science. To study the relationship between the response variable $Y^{*}$ and a set of covariates $X$, linear regression methods have been widely applied, which assumes that $E(Y^{*}|X)=X^{T}\beta$ and $\beta$ is an unknown parameter vector \cite{seber2012linear,neter1996applied,huang2020bi,huang2019bi,huang2019decomposition}. 
However, in practice, the response variable $Y^{*}$ may not be fully observed because of data issues, such as censoring or detection limit, which is common in economic and biomedical research\cite{tobin:1958,ORKIN2020e389,huang2020high}. One example in medical research is that in HIV clinical studies, the HIV viral load that usually serves as the primary endpoint in HIV clinical trials, has a lower limit of detection of 50 or 20 copies/ml \cite{ORKIN2020e389,wohl2019bictegravir,erlandson2021weight,acosta2021three,orkin2019long,brar20201028,castagna2020548}. And another example in ecomomic research is that in household economics survey, household expenditures on various categories of goods vary with household income, and often censored at some lower limit\cite{tobin:1958}. 
In such situation, instead using traditional regression approach, certain statistical techiniques would need to be employed to adjust for data issue.
In the literature, for regression with a response variable having data issue such as censoring or detection limit, researchers proposed Tobit regression models and have been extensively studied \cite{tobin:1958,amemiya1973regression,amemiya:1984,huang2020estimation}. In Tobit models, the linearity assumption between latent true response variable $Y^{*}$ and $X$ is maintained as in linear regression models, and the censoring or detection limit problem has also been taken account  \cite{tobin:1958,amemiya1973regression,amemiya:1984}, i.e.
\begin{eqnarray}
\label{eq:00}
Y^{*} &=&X^{T}\beta  +\epsilon,\\ \nonumber
Y &=& \max(Y^*,c) \nonumber,
\end{eqnarray}
where $c$ is some fixed detection limit, $\epsilon$ follows normal distribution with mean zero and some unknown variance $\sigma$, and $Y$ is the observable value of latent true response variable $Y^*$. Likelihood approaches could be used to obtain statistical consistent estimation for (\ref{eq:00}) \cite{tobin:1958,amemiya1973regression,amemiya:1984}.
However, it is worthy noting that (\ref{eq:00}) may be subject to linear shape restrictions. In the literature for regression analysis, additive models have been broadly acknowledged as an extension for linear models to increase model flexibility and have been widely investigated\cite{hastie1987generalized,hastie1990generalized}.
As a natural consequence, we consider extending Tobit models to Tobit aditive models by replacing the linear form $X^{T}\beta$ with additive components $\sum_{j=1}^{d} m_{j}(X_{j})$, where $m_{j}(\cdot)$ is an unspecified smooth function for a $d$ dimensional covariate $X$.  
Although there are a great number of studies in the literature for Tobit and additive models, in the literature, a direct hybrid of Tobit and additive models has not been investigated.

In this manuscript, we propose semi-parametric Tobit additive models, which is more flexible than traditional Tobit models and also takes censoring problem into consideration. 
The method is easy to implement with well-established optimization techniques, and can perform well even when sample size is relatively small.
The manuscript is organized as follows,
Section II introduces  Tobit additive models and the likelihood-based estimation procedure. Section III explores the performance of the method through numerical simulation studies.
Section IV states conclusions and future study directions.

\section{Model formulation and Methods}\label{Model and estimation methodology}

\subsection{Model Formulation}
\noindent
We assume the following model for the latent response,
\begin{eqnarray}
\label{eq:21}
&&Y^{*} =\sum_{j=1}^{d}m_{j}(X_{j})  +\epsilon,\\ \nonumber
&&\epsilon \sim N(0,\sigma^2) \nonumber,
\end{eqnarray}
where $X = (X_{1}, \ldots, X_{d})^{T}$ is a $d$-dimensional covariate, $m_j(\cdot)$'s are unspecified smooth functions which are standardized such that $E(m_{j}(X_{j}))=0$ \cite{hastie1987generalized,hastie1990generalized}.
The observed $Y_{i}$ is such that $Y_i=\max(Y_i^*,c)$, where $c$ is the known lower detection limit, that we will assume to be zero W.L.O.G. \cite{tobin:1958,amemiya1973regression}. 
Let $\delta_i=I(Y_i^*>0)$,
and we assume that $\epsilon_i$'s are independently and identically distributed (i.i.d.) from normal distribution with mean $0$ and unknown finite variance $\sigma^{2}$.




\subsection{Likelihood-Based Estimation Approach}\label{Estimation method}

The likelihood based estimation method combines the idea in estimation of additive models \cite{hastie1987generalized,buja1989linear} and  maximum likelihood estimation methods for Tobit models in the literature \cite{tobin:1958,amemiya1973regression}.  In the estimation of Tobit models (\ref{eq:00}), $\beta$ could be obtained by maximizing 
\begin{eqnarray}
\label{eq:a0}
L_{n}(\gamma) &=& \prod_{i=1}^n\left\{ \frac{\phi\left(\frac{Y_i-X^{T}\beta }{\sigma} \right)}{\sigma}\right\}^{\delta_i}
\left\{1-\Phi\left(\frac{X^{T}\beta}{\sigma}\right)\right\}^{(1-\delta_i)}.\nonumber
\end{eqnarray}

Naturally, by replacing $X^{T}\beta$ by $m(X)= \sum_{j=1}^{d}m_{j}(X_{j})$, the likelihood function for our model is thus given by 

\begin{eqnarray}
\label{eq:a1}
L_{n}(\gamma) &=& \prod_{i=1}^n\left\{ \frac{\phi\left(\frac{Y_i-m(X) }{\sigma} \right)}{\sigma}\right\}^{\delta_i}\nonumber
\\
&&\times\left\{1-\Phi\left(\frac{m(X)}{\sigma}\right)\right\}^{(1-\delta_i)}.
\end{eqnarray}

Then we approximate each $m_j(\cdot)$ using B-spline \cite{hastie1987generalized}. Suppose $\kappa \in \mathbb{N}^{+}$ B-spline basis functions would be used to approximate each nonparametric additive components, for notation simplicity, we define
basis functions and the associated coefficients as
\begin{eqnarray}
B_{\kappa}(X)&=& \Big\{b_{1}(X_{1}),\ldots,b_{\kappa}(X_{1}),b_{1}(X_{2}),\\
&&\ldots,b_{\kappa}(X_{2}),\ldots,b_{1}(X_{d}),\ldots,b_{\kappa}(X_{d}) \Big\}^{T},\notag\\
\theta_{\kappa}&=&\Big(\theta_{11}, \ldots,\theta_{1\kappa}, \theta_{21}, \ldots,\theta_{2\kappa},\ldots,\theta_{d1}, \ldots,\theta_{d\kappa}\Big)^{T}.\notag
\end{eqnarray}
Let $\hat{m}(X)=B_{\kappa}(X)^{T}\theta_{\kappa}$, for $X_{j} \in [0,1]$, $~j=1,\ldots,d$, and
we find $\hat{\theta}_{\kappa}$ and $\hat{\sigma}$
by maximizing
\begin{eqnarray}
\label{eq:a1}
L_{n}(\theta_{\kappa},\sigma) 
&=&\prod_{i=1}^n\left\{ \frac{\phi\left(\frac{Y_i-B_{\kappa}(X)^{T}\theta_{\kappa} }{\sigma} \right)}{\sigma}\right\}^{\delta_i} \nonumber\\ 
&&\times\left\{1-\Phi\left(\frac{B_{\kappa}(X)^{T}\theta_{\kappa}}{\sigma}\right)\right\}^{(1-\delta_i)},
\end{eqnarray}
where $\phi$ and $\Phi$ are the standard normal density and distribution functions, respectively. Correspondingly, it is also equivalent to maximimize the log-likelihood function  given below
\begin{eqnarray}
\label{eq:a2}
\log(L_{n}(\theta_{\kappa},\sigma))
&=&\sum_{i=1}^n \delta_i \{-\log{\sigma} - \frac{(Y_i - B_{\kappa}(X)^{T}\theta_{\kappa} )^{2}}{2\sigma^2}\} \nonumber\\
&&+ \sum_{i=1}^n (1-\delta_i)\log\{1 \nonumber\\
&&- \Phi(\frac{B_{\kappa}(X)^{T}\theta_{\kappa}}{\sigma})\}.
\end{eqnarray}
Numerically, well-established optimization methods like Newton-Raphson or BFGS algorithm can be used to solve (\ref{eq:a1}) and (\ref{eq:a2}). 
In addition, to select the tuning parameter $\kappa$ that used in the B-spline basis functions, we borrow the idea of cross-validation, and propose a 5-fold cross-validation likelihood score function (for parameter tuning) as follows:
$$CV(\kappa)=\sum_{l=1}^{5}\log(L_{n}(\theta_{\kappa},\sigma))^{-l},$$
where $\log(L_{n}(\theta_{\kappa},\sigma))^{-l}$ is the log likelihood with the $l$th folder removed and using tuning parameter $\kappa$. We maximize this function over a grid of $\kappa$ values and choose the $\kappa$ which yields the maximum cross-validation score.

\section{NUMERICAL EXPERIMENTS}

In this section, we investigate the finite sample performance of our proposed model and methods by Monte Carlo simulations. We generate $50$ replicates, each consisting of $n$ observations from the following model :
\begin{eqnarray}
\label{eq:01}
Y^{*} &=&\sum_{j=1}^{2}m_{j}(X_{j}) + 0.2 \epsilon,
\end{eqnarray}
where $n$ is chosen to be $80$ or $160$, and $\epsilon$ follows a standard normal distribution.
Both  $X_{1}$ and $X_{2}$ are generated independently and following an uniform distribution on $[0,1]$.
The components of the additive model (\ref{eq:01}) are:  $m_{1}(v)=v-0.5 $ and  $m_{2}(v)=(s-0.5)^2-1/12$, where one component is linear and the other one is nonlinear.
The observed response variable $Y_i = \max(Y^{*}_{i},c)$, where $c$ is determined by censoring proportions (Cens): Cen=5\%, Cen=15\% and Cen=30\%.

To implement our likelihood-based estimator, the basis functions used  in the estimation are chosen to be cubic  B-splines  with $1$ inner knots decided by cross-validation, the selection of cubic B-spline is based on some previous numerical studies and its desired theoretical property \cite{horowitz2004nonparametric,horowitz2011oracle}.
Since the Tobit model can only work with linear relationship, it is by theory can not work with nonlinear relationship, we compare the performance of our method with nonparametric estimation method (NP) for additive models \cite{doi:10.1080/10485252.2018.1537441}.
Table  1 summarizes the  IIMSE (integrated mean square error) of  $\hat{m}_{1}(\cdot)$ and $\hat{m}_{2}(\cdot)$ at $50$ equally-spaced grid points over $[0,1]$. From Table  1, it can be seen that in general, the IMSE's of Tobit Additive method are much smaller than those of the NP method. 
For both methods, the IMSE's increase when Cen's increase, and the IMSE's will decrease with the increase of the sample size.


\begin{table}[htbp]\label{tab:b1}
	\begin{center}\tabcolsep 0.5mm\caption{
			$10^4\times$IMSE of $\hat{m}_{1}(\cdot)$ and $\hat{m}_{2}(\cdot)$  at $50$ equally-spaced grid points  over $[0,1]$.}
		\begin{tabular}{c c  c c c  c c c c c}
			\hline\hline
			$n$& Cen & \multicolumn{2}{c}{Tobit Additive} & \multicolumn{2}{c}{NP}&\\
			&  &  IMSE($m_1$) & IMSE($m_2$)  & IMSE($m_1$)  & IMSE($m_2$) \\ \hline
			
			80	& 5\% & 26.9  &26.0 &60.9 & 41.9 \\
			& 15\% & 35.6 & 27.5 & 79.6 & 42.2 \\
			& 30\% & 67.9 &31.9 & 107.7 & 44.3  \\
			
			160 & 5\% & 12.5  &10.5 &36.5 & 18.4 \\
			& 15\% &13.3 &10.8 &42.6& 19.1 \\
			& 30\% & 19.0 & 11.9 & 72.7 & 20.9 \\
			\hline
			\hline
		\end{tabular}
	\end{center}
\end{table}

\begin{figure}[htbp]
	\begin{center}
		\includegraphics[width=0.5\textwidth,height=0.65\textheight]{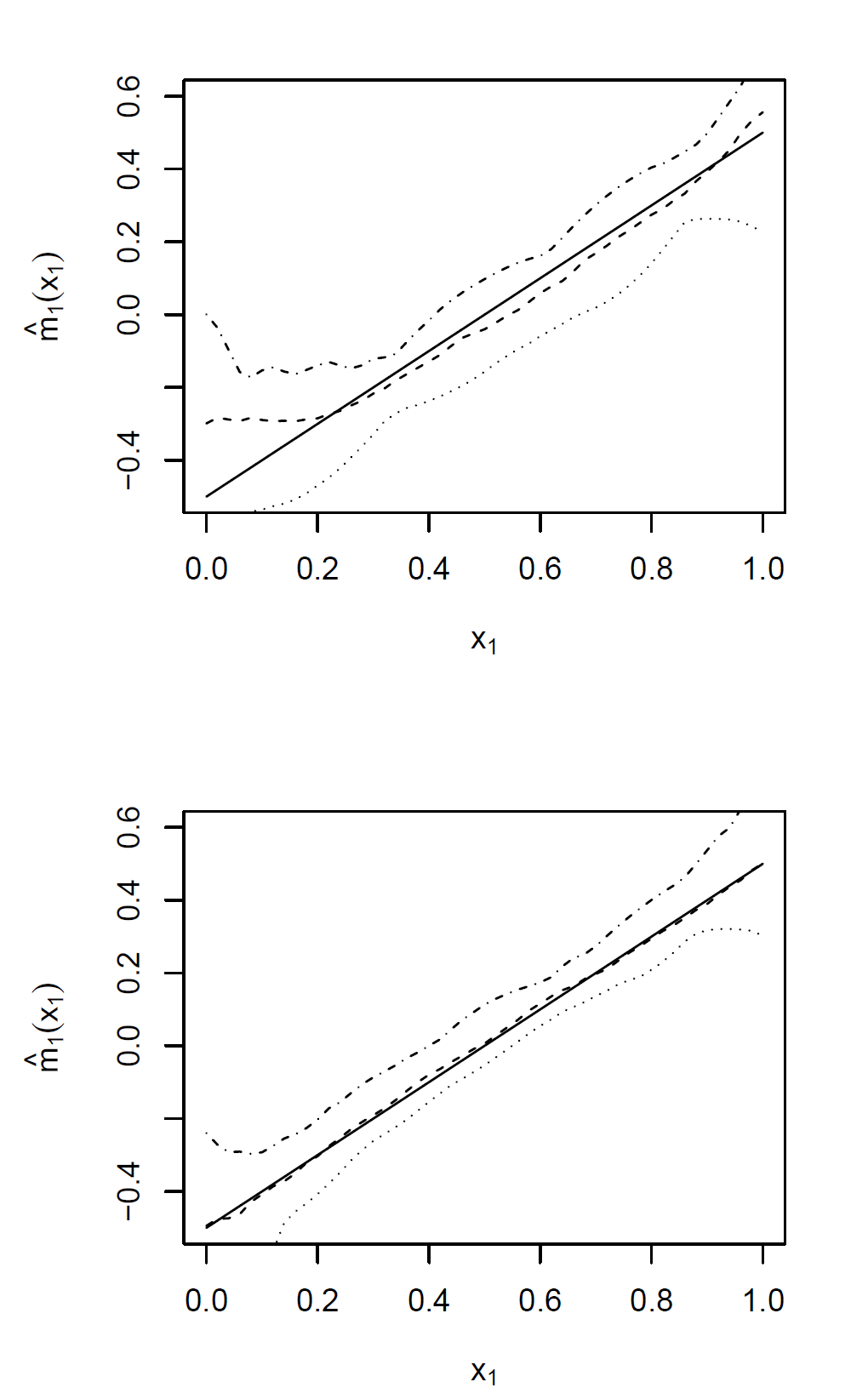}
		\caption{For linear function:the true curves (solid lines), estimated median curves (dashed  lines),and associated $95\%$ confidence bands (dotted lines) for $m_1(\cdot)$, using the Tobit Additive method, the NP method (from bottom to upper) when cen=$30\%$. \label{fig:41}}
	\end{center}
\end{figure}

\begin{figure}[htbp]
	\begin{center}
		\includegraphics[width=0.5\textwidth,height=0.65\textheight]{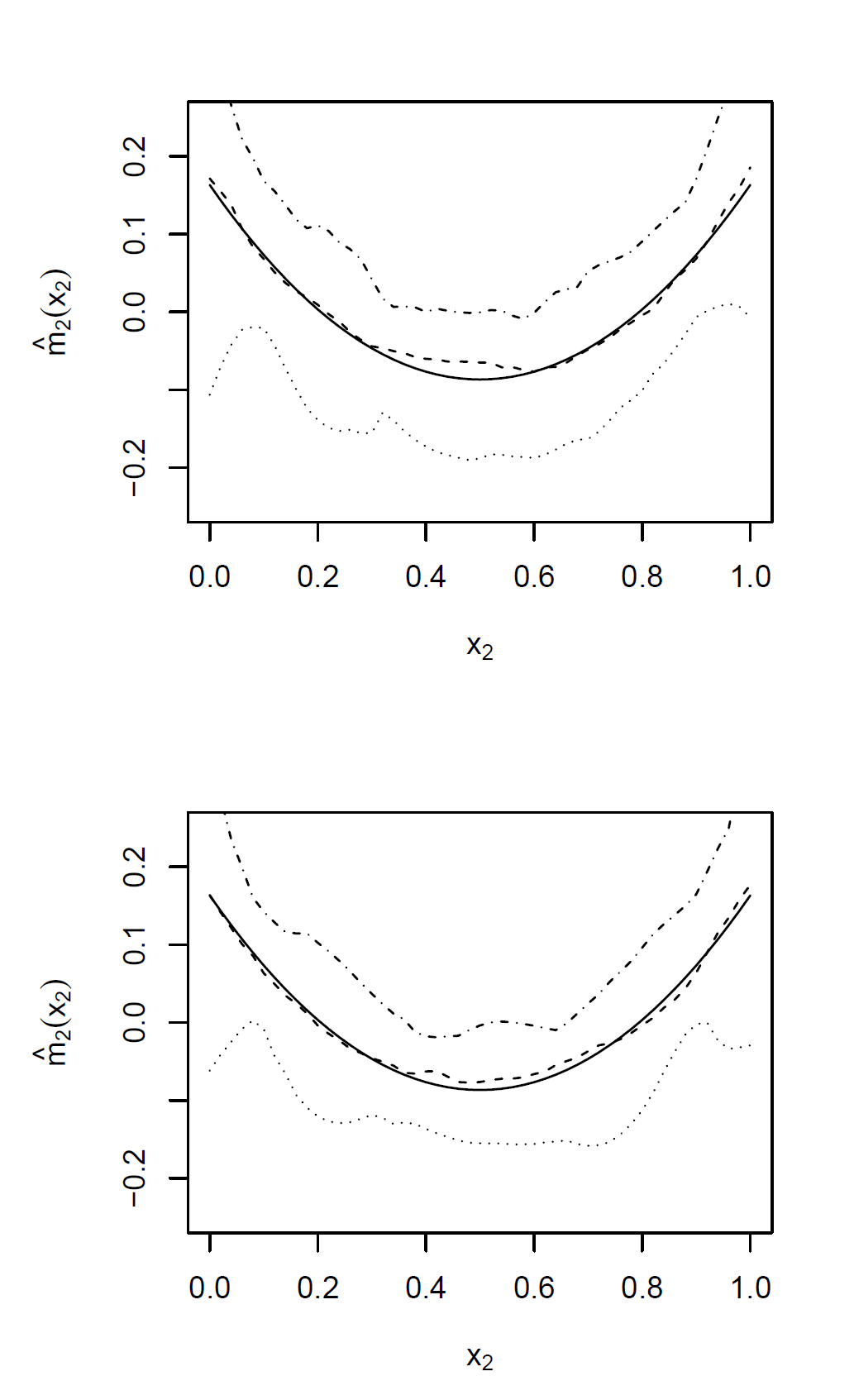}
		\caption{For quadratic function: the true curves (solid lines), estimated median curves (dashed  lines),and associated $95\%$ confidence bands (dotted lines) for $m_2(\cdot)$, using the Tobit Additive method, the NP method (from bottom to upper) when cen=$30\%$. \label{fig:42}}
	\end{center}
\end{figure}

Figures (1) and (2) present the point-wise median curves of the estimated component functions $\hat{m}_{1}(\cdot)$ and $\hat{m}_{2}(\cdot)$ at the selected grid points when $n=80$ and Cens $=0.3$, as well as the $95\%$ confidence bands.
The discrepancy between the true curves and fitted median curves provide a measure for the bias of the estimators, and the 2.5\% and 97.5\% lines demonstrate the variability of the estimators.
From Figures  1, it can be seen that the fitted values for both methods are close to the true values, and the confidence bands in general covers the true curves, while the Tobit Additive method can produce a better point estimation and narrow confidence bands compared with the NP approach.

\section{Conclusions}
In this research article, a novel semi-parametric Tobit additive regression model was proposed to account for censoring or limit detection in regression analysis and to extend flexibility of Tobit models. The method is straight forward and easy to implement, and perform well in numerical experiments. In the future, the research direction would be in theoretical perspective, in which asymptotic properties for the estimators could be further investigated.

\bibliographystyle{IEEEtran}
\bibliography{tobit_refall_new}{}

\end{document}